\documentstyle[twocolumn,psfig]{mn}

\def\ltsima{$\; \buildrel < \over \sim \;$}
\def\lsim{\lower.5ex\hbox{\ltsima}}
\def\gtsima{$\; \buildrel > \over \sim \;$}
\def\gsim{\lower.5ex\hbox{\gtsima}}

\begin{document}
\title[Evidence of self-interacting cold dark matter]
{Evidence of self-interacting cold dark matter from
galactic to galaxy cluster scales}

\author[Firmani, D'Onghia, Avila-Reese, Chincarini, Hern\'{a}ndez] 
{Firmani C.$^{1,3}$, D'Onghia E.$^{2}$, 
 Avila-Reese V.$^{3}$, Chincarini G.$^{1,4}$ $\&$ 
 Hern\'{a}ndez X.$^{5}$\\
 $^{1}$ Osservatorio Astronomico di Brera, 
 via E. Bianchi 46, 23807 Merate (LC), Italy\\
$^{2}$ Universit\'{a} degli Studi di Milano, via Celoria 16, 
 20100 Milano, Italy\\
$^{3}$ Instituto de Astronom{\'\i}a, UNAM, A.P. 70-264, 04510 
M\'{e}xico D.F.,
 M\'{e}xico \\
$^{4}$ Universit\'{a} degli Studi di Milano-Bicocca, Italy\\
$^{5}$ Osservatorio Astrofisico di Arcetri, Largo E. Fermi 5,
 50125 Firenze, Italy \\
E--mail: {\tt firmani@merate.mi.astro.it}, {\tt elena@merate.mi.astro.it}, 
{\tt avila@astroscu.unam.mx},\\ {\tt guido@merate.mi.astro.it}, 
{\tt xavier@arcetri.astro.it}
}

\date{\underline{Submitted to MNRAS, January 20, 2000}}
\maketitle

\begin{abstract}

Within the framework of the cold dark matter (CDM) 
cosmogony, a central cusp in the density 
profiles of virialized dark haloes is predicted. 
This prediction disagrees with the soft inner halo mass 
distribution inferred from observations of dwarf and low 
surface brightness galaxies, and some clusters of galaxies. 
By analysing data for some 
of these objects, we find that the halo central
density is nearly independent of the mass from galactic to galaxy
cluster scales with an average value of around
$0.02 \ M_{\odot}/pc^3$. We show that soft cores can be produced in the CDM
haloes by introducing a lower cut-off in the power
spectra of fluctuations and assuming high orbital thermal 
energies during halo formation. However, the scale invariance
of the halo central density is not reproduced in these cases. The 
introduction of self-interaction in the CDM particles offers the 
most attractive alternative to the core problem. We propose
 gravothermal expansion as a possible mechanism to 
produce soft cores in the CDM haloes with self-interacting
particles. A global thermodynamical equilibrium can explain the
central density scale invariance. We find  a minimum
 cross section capable of establishing isothermal cores in agreement with
the observed shallow cores. If $\sigma$ is the cross section, $m_x$
is the mass of the dark matter particle and $v$ is the halo velocity 
dispersion, then 
$\sigma /m_x \approx 4 \ 10^{-25} (100 \ km s^{-1}/v)$ $cm^2/GeV$.

\end{abstract}

\begin{keywords}
galaxies: formation - galaxies: haloes - clusters: haloes - 
           cosmology: theory - dark matter 
\end{keywords}

\section{Introduction}

Constraining the nature of dark matter is presently one of the most 
relevant problems in cosmology and particle physics. The 
current most popular scenarios for structure formation in 
the universe are based on the inflationary 
 CDM theory, according to which cosmic structures arise 
from small Gaussian density fluctuations composed
mostly of non-relativistic collisionless particles. Luminous 
galaxies are thought to form by 
gas cooling and condensing into the dark matter haloes 
which grow by gravitational accretion and merging in a hierarchical fashion.

The question on the inner density profiles of 
the virialized dark matter haloes is at present controversial.
In the last few years much observational and theoretical 
effort has been employed into investigating the inner structure 
of dark haloes. On galactic
scales, the rotation curves of dwarf galaxies offer a 
way to study the inner mass distribution of their dark haloes
 directly since these galaxies are dominated by dark matter. By
analysing the rotation curves of some near dwarf galaxies, 
Moore (1994), Flores $\&$ Primack (1994) and  Burkert (1995) have 
shown that the central mass distribution of their dark haloes is 
soft, i.e. the haloes have a constant density core. A similar result 
concerns low surface brightness galaxies (LSB, hereafter) (de Blok  
$\&$ McGaugh 1997) even though the uncertainty in  
the observational data is larger than in the case of dwarf spirals.
Hern\'{a}ndez $\&$ Gilmore (1998) showed that the observed
rotation curves of both LSB and normal large galaxies are
consistent with a fixed initial halo shape, characterized by a significant
soft core inner region.
On scales of clusters of galaxies, unfortunately there is not 
much information available. Recently, from strong gravitational 
lensing observations, Tyson, Kochanski, \& Dell'Antonio (1998) have
obtained an unprecedent high-resolution mass map for the cluster 
CL0024+1654, which has not a central cD galaxy, and found the 
existence of a soft core. Taken together these studies suggest 
the univesality of constant density cores across both large mass
scales and galactic types.

On the theoretical side, the structure of the CDM
haloes was studied over a wide range of masses by 
means of high-resolution N-body cosmological simulations 
(e.g., Navarro, Frenk, \& White 1997; NFW hereafter) 
and semi-analytical 
approaches (e.g., Avila-Reese, Firmani, \& Hern\'{a}ndez 1998).
 It was found that
the universal density profile firstly introduced by NFW describes very well 
the mass distribution of most of the CDM haloes. This profile is 
univocally determined by the mass, and in the centre diverges as 
$\rho \propto r^{-1}$ producing a cusp in the core. Recent high-resolution 
N-body simulations have shown that, as the numerical 
resolution is increased, the inner profiles result even steeper than
$r^{-1}$ (e.g., Moore et al. 1999b), making the CDM haloes
more cuspy than in the case of the NFW profile.

So far, the predicted inner density profile of the CDM haloes seems 
to be in conflict with the observations. Another potential 
difficulty for the CDM models was recently reported: the N-body 
simulations predict an overly large number of haloes within group-like 
systems compared to observations (Klypin et al. 1999; Moore et al. 1999a). 
In light of these difficulties, the current stance of the 
 hierarchical CDM-based 
scenario of structure formation remains somewhat confusing
because, in fact, this scenario successfully accounts for: the distribution 
of matter at large scales (Bahcall et al. 1999), the 
uniformity of the cosmic microwave radiation and its small temperature 
anisotropies, and the observationally inferred cosmological 
parameters. 

The aim of this letter is to analyse the halo core properties inferred
from observations which might suggest explanations of the origin as to the
soft halo cores and clarify the discrepancies that appear on small scales
with the hierarchical scenario of structure formation.
We investigate whether some modifications on the initial
conditions of this scenario
are able to improve the results with respect to the observations. 
We demonstrate that the introduction of self-interaction in the
CDM particles as was suggested by Spergel \& Steinhardt (1999) 
offers the most viable solution to the core problem in a context
that preserves the hierarchical CDM-based scenario.

\section{Halo central density  from observations}

We select from the literature dwarf and LSB galaxies with 
accurately measured rotation curves and clearly dominated
by dark matter. These restrictions considerably reduce 
ambiguities in the estimates of the dark matter mass distribution 
due to uncertain stellar mass-to-light ($M/L$) ratios and 
modifications of the original halo profile produced by the 
gravitational drag of baryons during disc formation. Hence 
the dark haloes of these galaxies can be rightly assumed almost ``virgin''. 
These constraints reduce the sample to six dwarf galaxies: 
DDO154 (Carignan  et al. 1998), DDO170 (Lake et al. 1990), 
DDO105 (Schramm 1992, quoted by Moore 1994), 
NGC3109 (Jobin et al. 1990), IC2574 (Martimbeau et al. 1994),
NGC5585 (C\'{o}te et al. 1991).
Six LSB galaxies are selected with
the same criterion from a published sample: 
F568-v1, F571-8, F574-1; F583-1, F583-4, UGC5999
(de Blok $\&$ McGaugh 1997).
The rotation curves measured for all these galaxies were used
by the different authors to estimate the halo parameters,
particularly the central density.

Our analysis also includes the density profile obtained for the 
cluster CL0024+1654 from a high resolution mass map derived using
strong lensing techniques (Tyson et al. 1998). 
Because of the lack of a massive cD galaxy in the core, this 
cluster can be assumed to be dark matter dominated at the centre. 
Two clusters of galaxies, CL1455+22 and CL0016+16, 
with evident shallow mass profiles in the inner regions obtained by
weak gravitational lensing studies (Smail et al. 1995)
have also been considered, even though the uncertainty of 
the observational data is larger in these cases.

In Figure 1 we plot a very suggestive result: for a broad range of
masses, the central density of the dark haloes is independent of
 mass (or circular velocity). Most dwarf galaxies 
(filled squares), LSB galaxies (open squares) and clusters 
(circles) indicate an average halo core density 
close to $\rho_c=0.02 \ M_{\odot}/pc^3$. 
The arrow shows a fiducial value derived from a published sample
of LSB galaxies (de Blok $\&$ McGaugh 1997).
The galaxy error bars
are based on the observational uncertainty, and when possible from
the range given by the maximum and minimum disc models.  
 The cluster error bars take into account the uncertainty
in observations and in a normalization factor of three in 
going from strong to weak lensing techniques (Wu et al. 1998).

This observational evidence makes the cosmological puzzle quite complex:
how can one explain the origin of soft halo cores with roughly the same 
central density over the entire mass range sampled?

\section{Shallow cores from collisionless cold dark matter}

As was discussed above, observations seem to show that the
inner density profile of the dark matter haloes is (i) shallow, and
(ii) with a central value independent
from the total halo mass (or maximum circular velocity $V_m$). 
These facts disagree with the 
predictions of the hierarchical CDM models. Now, we  investigate
some alternatives which might alleviate these difficulties within the 
cosmological context. For this we have performed a quantitative
study of the CDM halo profiles using a semi-numerical method 
(Avila-Reese, Firmani, \& Hern\'{a}ndez 1998) aimed at calculating
the collapse and virialization of spherically symmetric density 
fluctuations starting from an arbitrary mass aggregation history. 
Results obtained with this method are in excellent agreement
with those of the N-body simulations (see Avila-Reese et al. 1999;
Firmani \& Avila-Reese 2000). The
method is based on a generalization of the secondary infall model
where non-radial motions and adiabatic invariance are 
taken into account. The only free parameter is the orbital 
parameter of particles  (the perihelion 
to aphelion ratio) which regulates the thermal orbital energy of the
system. This parameter is fixed independently of the halo mass
and is constant during halo formation. Cosmological
N-body simulations suggest $r_{\rm peri}/r_{\rm apo}\approx 0.2-0.3$ 
(see Ghigna et al. 1998).

Recently, Moore et al. (1999b) have simulated CDM haloes formed by
monolithic collapse with N-body simulations introducing for this
a lower cut-off at some wavelength in the power spectrum 
of fluctuations which suppresses substructures.
The result was that the steep inner density profile of the haloes persisted. 
We suggest that this result might be partially a consequence of the 
lack of thermal orbital energy. 
In a monolithic collapse scenario the 
thermal orbital energy plays a significant role in producing soft cores:
as $r_{\rm peri}/r_{\rm apo}$ increases a larger soft core is obtained.
The density profiles of our haloes obtained for a CDM model with 
a lower cut-off in the variance of the power spectrum and a non zero initial
thermal  content, present soft cores. However, these
models are unable to predict the observed central density trend 
shown in Figure 1 (Avila-Reese et al. 1998). In fact, the 
central density $\rho _c$ increases with $V_m$ in such a way
that if $\rho_c$ is reproduced at galactic scales, for the cluster 
scales, $\rho_c$ overshoots the observed value by more than an order of
magnitude. A hypothetical injection of
thermal energy to the dark matter at a specific time in the life of the
universe leads to a similar negative result.

An interesting way to 
produce soft halo cores in agreement with observations is to  
 simply truncate the hierarchical
halo mass aggregation histories at a given redshift towards the 
past. This may be done assuming that the halo mass fraction  
instantaneously collapses with some thermal energy (monolithic 
 thermal collapse),
while the rest of the mass is aggregated at the normal hierarchical rate.
 We have calculated the 
density profiles for haloes whose mass aggregation histories 
correspond to a hierarchical flat $\Lambda $CDM model 
($\Omega _m=0.3$, h=0.7, $\sigma _8=1$) from $z=5$ and 
$r_{peri}/r_{apo}=0.3$; before this 
epoch the hierarchical aggregation
was truncated. The results for this toy model are in good 
agreement with the observations: the haloes have a soft core,
the core densities are independent from the mass and have
a value similar to that what observational inferences indicate.
It is interesting to note that the most distant QSOs and galaxies 
are at redshifts $z\approx 5$.
Although the toy model presented here might look  attractive, it is
difficult to imagine a physical process capable of delaying the
collapse of the central parts of the CDM haloes until $z\approx 5$. 

\section{Shallow cores from self-interacting cold dark matter}

 Self-interacting dark 
matter has been proposed as a possible
solution for two potential conflicts of the hierarchical CDM
models (Spergel \& Steinhardt 1999; Hannestad 1999): 
the shallow core of the haloes and the dearth of dwarf galaxies 
in the Local Group. Astrophysical consequences of collisional dark
matter have been pointed out by Ostriker (1999).
It is easy to show that a configuration 
with the NFW density distribution is very far from thermal 
equilibrium: the inner velocity dispersion (temperature) has 
a positive gradient. Consequently, the presence of some  
self-interaction in the CDM particles introduces in the dark 
haloes a process of thermalization with heat transfer inwards, 
avoiding the formation of a cuspy profile. Heat capacity in the core  
is negative. This is a typical property of 
self-gravitating systems, like the interiors of the stars. 
For this reason, the heat transfer inwards cools the core exacerbating
even more the temperature gradient. The heat transfer inwards increases 
causing the core to expand and cool due to gravothermal instability,
leading to runaway core expansion. This physical
mechanism is the key point for core expansion if  
self-interaction is effective.  This process is similar to 
the post-collapse gravothermal instability well-known in dynamical 
studies of globular clusters (Bettwieser \& Sugimoto 1983) 
where the minimum central density is reached roughly 
after a thermalization time. 

The expansion of the core does not last forever. Since as the core 
expands the central density decreases, this would make the 
self-interaction less efficient and the core formation
mechanism a self-limiting process. Although
attractive, this mechanism is difficult to investigate because of our lack
of knowledge regarding the cross section of the self-interacting 
dark matter particles.
For this reason we start our analysis with a thermodynamical approach: 
we shall 
estimate the central density of CDM haloes assuming a  
thermodynamical equilibrium is reached due to strong self-interaction
of the CDM particles. The final result will be the formation in the
CDM halo of a central isothermal non-singular density profile 
established by competition
between 1) mass and energy hierarchical aggregation, and 2) 
the thermalization due to
self-interaction. The hierarchical mass and energy aggregation 
tends to stablish a NFW density profile (with the corresponding heat 
transfer inwards) while the self-interaction process tends to lead the 
system to a thermal equilibrium with the corresponding formation
of a shallow core. For a given mass, the halo formed by a hierarchical
mass aggregation identifies a gravitational binding energy (or $V_m$).
Using this mass and binding energy to rescale a thermodynamical equilibrium
configuration it is easy to find:\\
\begin{equation}
\rho_c = \alpha \ \frac{V_m^6}{M^2} \ M_{\odot}/pc^3
\end{equation}
where $V_m$ is in km/s, $M$ is the halo mass in M$_{\odot}$ 
and $\alpha$ is a constant given by the detailed shape of the final
equilibrium configuration. Since for the CDM haloes a tight relationship 
between their mass and circular velocity of the kind $M \propto V_m^n$ with
$n\approx 3.2$ is predicted (Avila-Reese et al. 1998,1999), eq. (1)
implies that $\rho _c$ is roughly invariant with respect to the mass
or $V_m$ as observations point out (Fig. 1). This strongly suggests 
 that indeed a thermalization process due to dark matter 
self-interaction is acting in the CDM haloes.

Unfortunately, there is not a single final thermal 
equilibrium configuration, and as Lynden-Bell \& Wood (1968)
pointed out, some of configurations
are even unstable. The King and Wooley configurations
are examples of systems that have reached thermal equilibrium. They are 
characterized by a form parameter that may be related  to the 
entropy of the system. A fiducial value for the central density  
of the CDM haloes with self-interaction may be estimated using a 
King or a Wooley profile at the state of maximum entropy 
(Lynden-Bell $\&$ Wood 1968). For these cases we derive respectively 
$\alpha = 1.3 \ 10^{9}$ (short-dashed line in Fig. 1) and $\alpha = 
2.6 \ 10^{9}$ (long-dashed line in Fig. 1) in the appropriate units.
The case of maximum entropy for a King profile corresponds to 
a value of the form parameter of $8.5$. A lower limit for 
the density may be roughly estimated from 
the dynamical evolution of globular clusters based on the 
Fokker-Planck approximation  (Spitzer $\&$ Thuan 1972),
starting from a uniform spherical distribution (this initial condition
will lead to a central density lower than the density reached by the
thermalization of a steep initial profile).
The rescaling for this model taken at the first thermal equilibrium
state gives us $\alpha = 1.7 \ 10^{8} $ (dotted curve in
Fig.1).

Global thermal equilibrium is reached when the self-interaction
cross section is sufficiently large in order for the characteristic time
scale of interactions across the overall halo to be shorter than the halo 
lifetime.
An opposite situation of minimum cross section is given when
self-interaction induces thermal equilibrium only in the region 
of the shallow core. In this case the central isothermal core 
appears surronded by a matter distribution characterised 
by a NFW profile. From the observational data it is possible now to
infer an estimate of the self-interaction cross section. If $n$ is
the dark particle number density, $\sigma$ the cross section and $v$
the dispersion velocity, assuming the collision time in the core 
 $\tau = 1/(n \ \sigma \ v)$  close to the Hubble time we obtain:\\
\begin{equation}
\frac{\sigma}{m_x} \approx 4 \ 10^{-25} 
 \Big( \frac{0.02 \ M_{\odot} \ pc^{-3}}{\rho_c} \Big)
 \Big( \frac{100 \ km \ s^{-1}}{v} \Big) \ cm^2/GeV
\end{equation} 
with $m_x$ the mass of the dark matter particle and $\rho_c$ the
central density.
It is interesting to point out that for velocity dispersions 
corresponding to galaxy clusters this value is close to the upper limit 
estimated by Miralda-Escud\'{e} (2000) from the 
observationally inferred ellipticity 
of the cluster MS21137-23.  

\section{Summary}

The discovery of a soft core in the cluster of galaxies CL0024+1254 
by strong gravitational lensing measurements and the rotation 
curves of dark-matter dominated dwarf and LSB galaxies indicate 
that dark matter haloes have shallow inner density profiles
from galactic to cluster scales. Studying in detail the observational
data available for these cosmic objects, we found that
the halo central density is nearly invariant with respect
to the mass from galactic to cluster sizes.

We investigated different mechanisms
and models for halo core formation within the hierarchical CDM 
scenario. We have shown that a lower cut-off at some wavelength
in the CDM power spectrum and the assumption of high particle 
orbital thermal energies  
produce soft cores in the haloes, but the invariance of $\rho_c$
with respect to the mass is not reproduced. A more viable
solution to the core problem is the introduction of self-interaction 
in the CDM particles. Being this the case, we proposed the 
gravothermal expansion as the mechanism responsible for the 
formation of soft cores in a hierarchical CDM scenario.  

Using a thermodynamical approach we have estimated the central 
density of haloes in the case of maximum efficiency for self-
interaction and found good agreement with the values inferred
from observations. The central density in this case scales with the halo 
mass and its maximum circular velocity as $\rho _c\propto V_m^6/M^2$. 
This result implies that $\rho _c$ is roughly constant because for the 
CDM haloes $M\propto V_m^n$ with $n\approx 3.2$. If thermal equilibrium
is restricted to the core, then the cross section given by eq.(2) may
be derived consistently with observations. The cases analysed here,
corresponding to a global and a local thermal equilibrium respectively,
represent two limiting cases between which dark matter
self-interaction may generate isothermal cores compatible with 
observations.
We exclude from our analysis the extreme case of a {\it very} strong
self-interaction which may lead the core to a gravothermal catastrophe
with a central density profile steeper than NFW. Such extreme 
assumption of large cross section may be immediately ruled out because
a singular isothermal core will be produced in contradiction with 
observations.

We stress the relevance confirming the existence of soft
cores with scale invariant densities would have. In particular, the 
construction of high-resolution mass maps with gravitational lensing
techniques for the inner regions of clusters is of great interest.

\section*{Acknowledgments}
ED thanks Fondazione CARIPLO  for financial support.


\begin{figure*}
\vspace{15cm}
\includegraphics{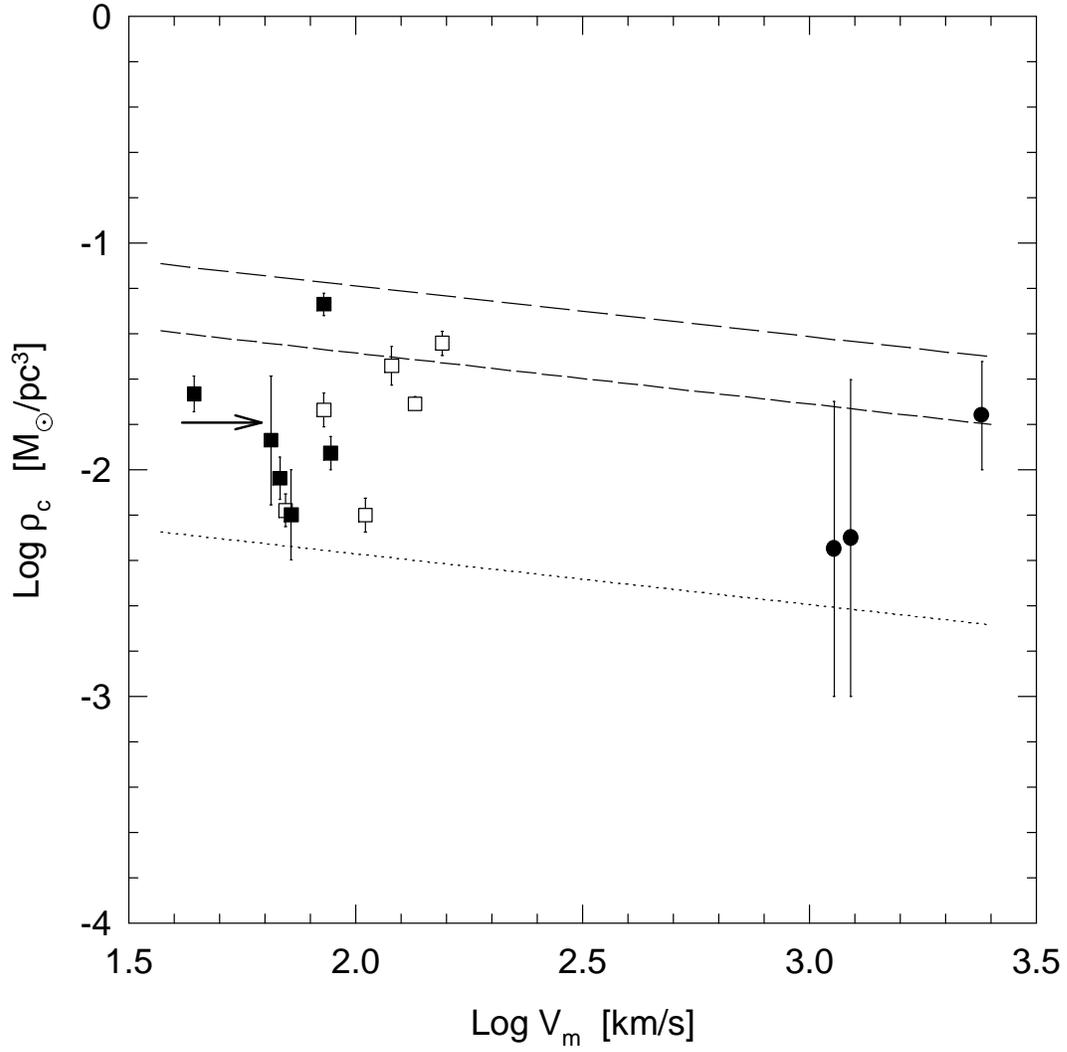}
\caption{ Halo central density  vs. maximum rotation velocity
for dwarf galaxies (filled squares), LSB galaxies (open squares) 
and galaxy clusters (filled circles).
 The plot shows a region predicted by gravothermal models 
 where a thermodynamical equilibrium is reached. The upper limit
 of the region (state of maximum entropy) is 
 estimated for a Wooley  (long-dashed line)
 and a King (short-dashed line) density profile. A  dynamical model 
  starting from a homogeneous sphere provides an estimate of the    
  lower limit for the halo 
 central density (dotted curve).}
\end{figure*}


\begin{thebibliography}{}

\bibitem []{}{Avila-Reese V., Firmani C., Hern\'{a}ndez X., 1998, ApJ, 505, 37}

\bibitem []{}{Avila-Reese V., Firmani C.,  Klypin A.,  Kravtsov A., 1999, 
 MNRAS, 309, 507}

\bibitem []{} {Bahcall N., Ostriker J.P., Perlmutter S., Steinhardt P.J.,
  1999, Science, 284, 1481}

\bibitem []{} {Bettwieser E., Sugimoto D., 1984, MNRAS, 208, 493}

\bibitem []{} {Burkert A., 1995, ApJ, 477,  L25}

\bibitem []{} {Carignan C., Burton C., 1998, ApJ, 506, 125}

\bibitem []{} {C\'{o}te S., Carignan C., Sancisi R., 1991, 
    AJ, 102, 904}

\bibitem []{} {de Blok W.J.G., McGaugh S. S., 1997, 
              MNRAS, 290, 533}

\bibitem []{} {Jobin M., Carignan C., 1990, AJ, 100, 648}

\bibitem []{} {Firmani C., Avila-Reese V., 2000, MNRAS, in press}

\bibitem []{} {Flores R., Primack J.R., 1994, ApJ, 427, L1}

\bibitem []{} {Ghigna S., Moore B., Governato F., Lake G., 
Quinn T., Stadel J., 1998, MNRAS, 300, 146}

\bibitem []{} {Hannestad S., 1999, preprint (astro-ph/9912558)}

\bibitem []{} {Hern\'{a}ndez X., Gilmore G., 1998, MNRAS, 294, 595}

\bibitem []{} {Klypin A., Kravtsov A.V., Valenzuela O., Prada
F., 1999, ApJ, 522, 82}

\bibitem []{} {Lake G., Schommer R.A., van Gorkom J.H., 1990, 
      AJ, 99, 547}

\bibitem []{} {Lynden-Bell D., Wood R., 1968, 
   MNRAS, 138, 495}

\bibitem []{} {Martimbeau N., Carignan C., Roy, J.R., 1994, 
    AJ, 107, 543}

\bibitem []{} {Miralda-Escud\'{e} J., 2000, ApJ, submitted (astro-ph/0002050)}

\bibitem []{} {Moore B., 1994, Nature, 370, 629 }

\bibitem []{} {Moore B., Ghigna S., Governato F., Lake G., 
Quinn T., Stadel J., 1999a, ApJ, 524, L19}

\bibitem []{} {Moore B., Quinn T., Governato F., Stadel J., Lake G., 
     1999b, MNRAS, submitted (astro-ph/9903164)}

\bibitem []{} {Navarro J., Frenk C.S., White S.D.M., 1997, ApJ, 490, 493}

\bibitem []{} {Ostriker J.P, 1999, preprint (astro-ph/9912548)}

\bibitem []{} {Smail I., Ellis R., Fitchett M.J., Edge A.C., 1995,
  MNRAS, 273, 277}

\bibitem []{} {Schramm D.N., 1992, Nucl. Phys. B. proc suppl. 28A, 243}

\bibitem []{} {Spergel D.N., Steinhardt P.J., 1999, preprint 
(astro-ph/9909386)}

\bibitem []{} {Spitzer L., Thuan T. 1972, AJ,  175, 31}

\bibitem []{} {Tyson J.A., Kochanski G.P., Dell'Antonio I.P., 
    1998, ApJ, 498, L107}

\bibitem []{} {Wu X.P., Chiueh T., Fang L.Z., Xue Y.J., 1998, 
       MNRAS, 301, 861} 

\end{thebibliography}
\end{document}